\documentclass[letterpaper,twocolumn,10pt]{article}
\usepackage{usenix,epsfig,endnotes}
\usepackage{epstopdf}
\usepackage[hyphens]{url}
\usepackage{graphicx}
\usepackage{color}
\begin{document}

\date{}

\title{\Large \bf Elasticizing Linux via Joint Disaggregation of Memory and Computation}

\author{
	Ehab Ababneh, Zaid Al-Ali, Sangtae Ha, Richard Han, Eric Keller\\
Department of Computer Science, University of Colorado Boulder
} 

\maketitle

\thispagestyle{empty}

\subsection*{Abstract}


In this paper, we propose a set of operating system primitives
which provides a scaling abstraction to cloud applications in which they can transparently be enabled to
support scaled execution across multiple physical nodes as resource needs go beyond that available on
a single machine.  These primitives include \emph{stretch}, to extend the address space of an application to
a new node, \emph{push} and \emph{pull}, to move pages between nodes as needed for execution and optimization,
and \emph{jump}, to transfer execution in a very lightweight manner between nodes.  
This joint disaggregation of memory and computing allows for
transparent elasticity, improving an application's performance by capitalizing on the underlying dynamic 
infrastructure without needing an application re-write.  We have implemented these primitives in a Linux 2.6 kernel,
collectively calling the extended operating system, ElasticOS.  Our evaluation across a variety of algorithms shows
up to 10x improvement in performance over standard network swap.

\section{Introduction}

We are in the midst of a significant transition in computing, where we are consuming infrastructure
rather than building it.  This means that applications have the power of a dynamic infrastructure underlying them,
but many applications struggle to leverage that flexibility.  In this paper, we propose supporting this at the
operating system level with new primitives to support scaling.  


To gain some context on the challenge with scaling, we first discuss how it is predominantly handled today.
The most straight forward option, which required no changes to applications, is to simply get a bigger (virtual) 
machine as load for an application increases.
Cloud providers, such as Amazon\cite{amazon}, offer a wide range of machine sizes which cost anywhere from
less than a penny per hour to a few dollars per hour.  For cost efficiency, companies wish to use 
the ``right'' size machine, which might change over time.  But, transitioning from one VM size to another 
can pose challenges.  In some cases, we can take snapshots (e.g., with CRIU~\cite{CRIU}) and migrate the application to a 
bigger/smaller VM, but this can be disruptive, and the management of the application needs scripts and
other infrastructure to trigger scaling.

An alternative is to re-write the applications with scaling in mind.  To leverage the scaling, commonly
applications are built around frameworks such as Hadoop\cite{hadoop}, Apache Spark\cite{spark}, MPI\cite{mpi} or PGAS \cite{pgas}. These frameworks are designed with the
flexibility of being able to execute tasks on a varying amount of distributed resources available.  
The problem here is two fold.  First, to leverage this, the application needs to be built for this -- a
significant challenge (requiring a re-write) for any existing application, and forcing application developers
to evaluate and become fluent in the latest frameworks and potentially adapt the application as the frameworks
change.  Second, and perhaps more challenging, is that not every application fits into one of these frameworks.

Another approach to scaling is to replicate VMs/containers when an application becomes popular and requires more resources.  This too introduces burdens on the programmer in order to synchronize shared data and state across
multiple replicas, as well as to script their applications to spawn/delete replicas depending on load.


In short, in each case, \emph{the burden of scaling is placed on programmers}.
We argue that developers shouldn't need to be experts in cloud management and 
other frameworks, in addition to also needing to be fluent in programming and their application domain.  Instead,
the operating system should provide more support.  Broadly speaking, the job of an operating system is to make the life
of an application developer easier (through abstraction).  A modern OS provides virtual memory abstractions, so developers
do not have to coordinate memory use among applications, network socket abstractions, so developers can send messages
without needing to be intimately familiar with the underlying network protocols, and many other abstractions (file system,
device, multi-tasking) all to support developers.\emph{We propose that scaling should be an OS abstraction}.


\textbf{Related Work:}
We are not the first to propose that operating systems should support scaling.   Scaling of memory approaches are popular and include efforts such as RAMCloud~\cite{RamCloud}, which requires refactoring in user space to utilize its memory scaling capabilities.  An early approach to sharing memory called DSM~\cite{DSM, mether,KaiLimemory,apollo,treadmarks,farm} suffered from scaling issues, but more recently disaggregation-based approaches towards memory have emerged that are centered around transparent scaling of memory behind the swap interface, such as NSwap, Infiniswap, X-Swap and Memx~\cite{Nswap, Infiniswap, X-Swap,memx}.
Scaling of computation approaches include process migration to machines with more resources~\cite{CRIU, BLCR, MOSIX,smith1988survey,milojivcic2000process,stellner1996cocheck}, in addition to the scaling frameworks and replication methods mentioned previously.  Approaches to accelerate process migration~\cite{fast96,Kerrighed} have been proposed to hide the latency of migration by copying most of the process state in the background and only copying a small delta to the new machine after halting the process.
Single system image (SSI) OSs such as Kerrighed, MOSIX, Sprite and Amoeba ~\cite{Kerrighed,MOSIX,sprite-migrate,amoeba} 
have been created to support operation across a distributed cluster of machines.  These approaches typically employ a process migration model to move computation around cluster nodes and require applications to be recompiled for these specialized OSs.

These prior efforts in OS scaling suffer from a variety of limitations.  Network swap-based approaches, while being a step in the right direction of disaggregation in the data center, miss the opportunity to exploit \emph{joint disaggregation} of computation and memory for improved performance.  Execution is typically assumed to be pinned on one machine, while memory pages are swapped back and forth remotely across the network.  This can result in excessive swapping of pages over the network.  In these cases, movement of computation to a remote machine towards a cluster of locality stored in the remote machine's memory would result in substantially faster execution and lower network overhead, as we will show later.

\begin{figure}
	\centering
	\includegraphics[width=0.4\textwidth]{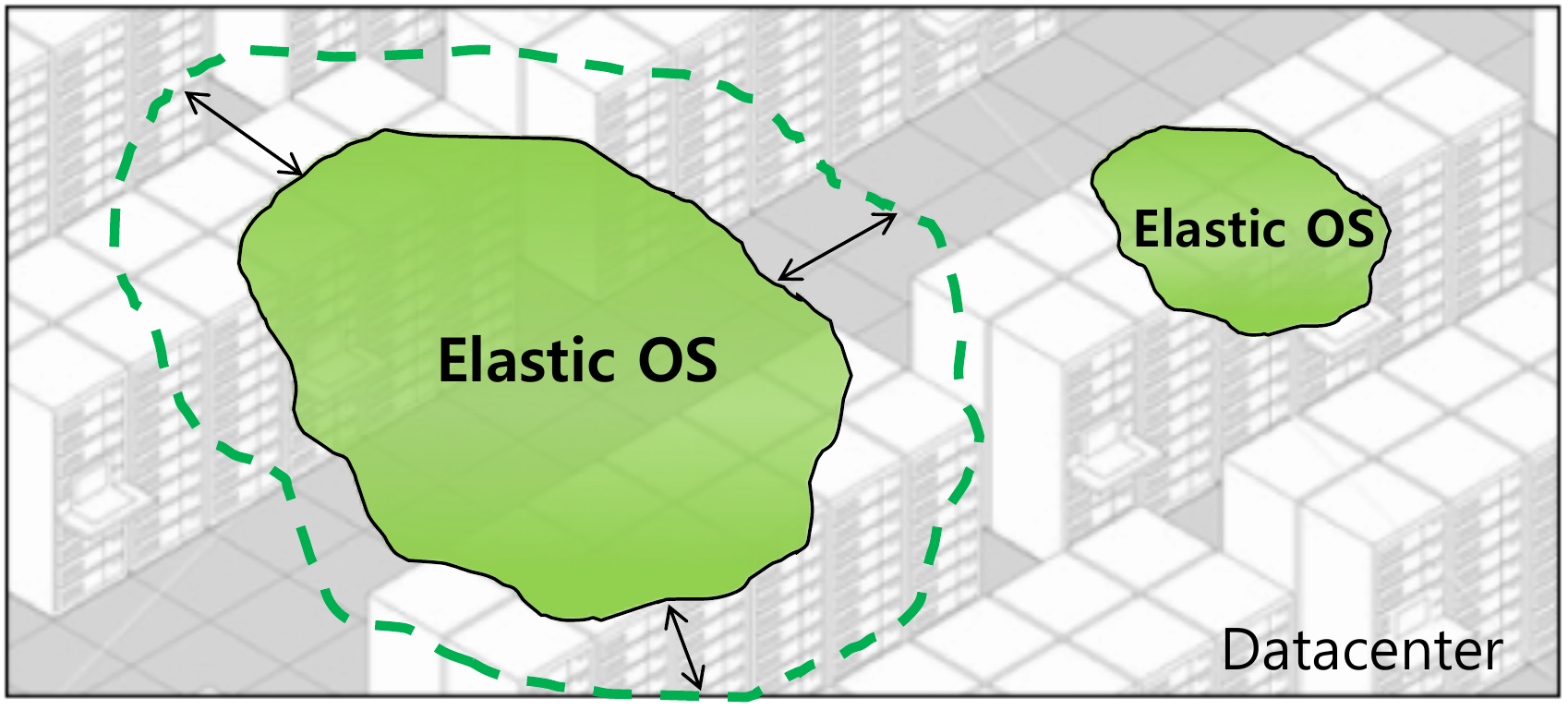}
	\caption{ElasticOS Vision for Cloud Data Centers.}
	\label{fig:Vision}
\end{figure}

Combining current network swap approaches with existing process migration techniques to alleviate excessive network swapping overhead would suffer two major limitations.  First, each decision to move computation would incur the overhead of copying the entire address space.  This is a significant amount of overhead to impose on the network.  Second, even with accelerated process migration, there is a substantial delay between the time the decision is made to migrate and when that is completed, at which time the conditions that triggered the original migration decision may be obsolete due to the length of time needed to copy all the state.



\begin{figure*}[th]
	\centering
	\includegraphics[scale=0.50]{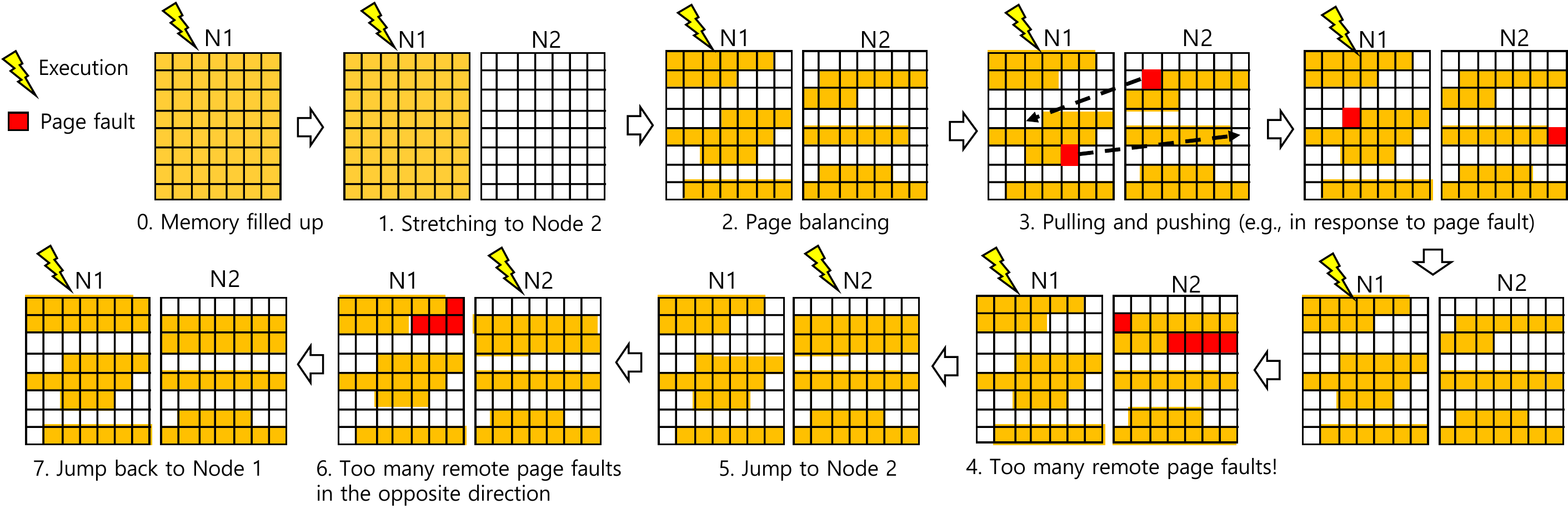}
	\label{fig:abstractions1}
	\vspace{-0.2in}
	\caption{Illustration of ElasticOS abstractions.  Each box labeled with a number above is a compute node,
		with the shaded boxes within represent individual pages.  Starting with execution on a single 
		machine in (0), when memory nears being filled, we stretch to two nodes in (1) and balance the pages in (2).  We then push and 
		pull pages in (3), with the red shaded pages going from node 1 to 2 (push) and from node 2 to 1 (pull).  Finally, in (4) and (6) we 
		are seeing too many page faults (resulting in pull), so decide to jump from node 1 to 2 in (5) and from node 2 to 1 in (7), respectively.}
\end{figure*}

\textbf{Introducing ElasticOS:}
In response to these shortcomings, we introduce four primitives to realize the scaling OS 
abstraction -- \emph{stretch}, \emph{jump}, \emph{push}, and \emph{pull}.  These scaling abstractions are designed to be transparent, efficient, and practically useful.  Our approach is inspired by an early work that
hypothesized elasticizing operating systems as a hot research topic, but did not build a working implementation
of the proposed concept~\cite{hotos13eos}.  \emph{Stretch} is used when an application becomes
overloaded (e.g., a lot of thrashing to disk is occurring), so the operating system \emph{stretches}
the application's address space to another machine -- extending the amount of memory available to the application.  
Push and pull allow memory pages to be transferred between 
machines which the application has been stretched to, whether proactively to optimize placement, or reactively
to make it so the data is available where it is needed.  
\emph{Jump} allows program execution to transfer to a machine which the application has been stretched to.  Unlike heavyweight process migration, our jump primitive is a lightweight transfer of execution that only copies the small amount of state needed to begin execution immediately on the remote machine, such as register state and the top of the stack.  Any additional state that is needed is faulted in using pulls from the rest of the distributed address space.  Having both jumping and push/pull allows for the OS to choose between moving the data to be where
the execution needs it, and moving the execution to be where the data is.  This supports the 
natural, but not necessarily perfect locality that exists in applications. 

To demonstrate the feasibility of this scaling approach, we extended the Linux kernel with these four primitives, and call the
extended Linux, ElasticOS.  
Figure~\ref{fig:Vision} provides a high level view of ElasticOS. 
We see that an instance of ElasticOS is capable of spanning a
number of nodes in the data center, and that the number of spanned nodes can 
elastically scale up or down depending upon application demand.  The application is executed
within ElasticOS, and the scaling primitives are used to support this execution across a distributed collection
of resources.

To demonstrate the desirability of these four primitives, 
we evaluated a set of applications with large memory footprints and compared against network swap, which
supports the pull and push primitives, and itself has shown performance improvements of being able to 
scale memory resources transparently across multiple machines.  We illustrate the additional
benefit of also transparently scaling computing resources across multiple machines, forming 
a system with joint disaggregation of memory and computation.
Our evaluation shows up to 10x speedup over network swap, as well as a reduction of 
network transfer between 2x and 5x.
        
In summary, we make the following contributions.
\begin{itemize}
\item Introduce scaling as a new OS abstraction, specifically with four primitives: stretch, push, pull, and jump. 
\item Provide an architecture and implementation of these abstractions in Linux. 
\item Demonstrate through an evaluation on Emulab servers that ElasticOS achieves up to 10x speed up over network swap across a range of applications, and up to 5X reduction in network overhead.
\end{itemize}

\if{false}

\begin{figure}
	\includegraphics[width=\linewidth]{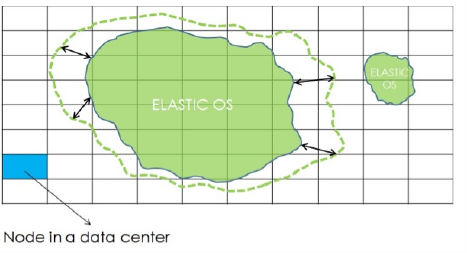}
	\caption{Vision.}
	\label{fig:Vision}
\end{figure}

In the past few years cloud computing has been gaining popularity due to its ability to help organizations scale up/down their compute infrastructure to meet the requirements of dynamically changing workload characteristics. Before that, enterprises built their own data centers and had to over-provision in order to keep up with varying compute power needs. By sharing a pool of resources, cloud computing created the illusion of unlimited resources. To understand the importance of this emerging technology consider an enterprise that offers a web based service housed in a classical data center. It is very crucial for the service to handle varying volume sizes, forcing the enterprise to over-provision when planning for capacity to meet business needs. This over-provisioning renders the data center resources under utilized during off-peak hours, days, or months. In contrast, consider the same service running in a cloud computing environment. It can respond to fluctuations of service demand by either one of two ways. First, it can scale horizontally by spinning up additional compute nodes to handle larger volumes during peak demand hours, and then, scale down by taking those additional nodes out of service and shut them down during off-peak hours. Second, it can scale up the service by migrating to a larger compute node during those peak hours, and scale down again by migrating back to a smaller one during off-peak ones. Certainly, this flexibility is made possible by the cloud’s promise of shared resources permitting on-demand compute power availability, which helps customers maximize their return on investment. \\

Cloud computing offers the ability to use more compute power on demand, but it is up to the cloud application to actually benefit from this offering. And what can limit its ability to do so is its architecture. In the web service application mentioned above, one approach to achieve dynamic scaling is that more web servers can be run in newly spawned nodes and then load balancers would redistribute the workload among all active servers. That may not be the case, however, for legacy applications, such as simulation software, or large graph analysis. For elasticizing such classes of applications, major recoding efforts and reconfigurations are necessary. Distributed systems research produced several works that seek to address this problem. Previous
approaches such as distributed shared memory (DSM) \cite{Farm,munin,KaiLimemory,mether,apollo}, MapReduce \cite{mapreduce}, message passing interface (MPI) \cite{mpi}, partitioned global address space (PGAS) \cite{de2015partitioned}, and remote paging based approaches \cite{memx,Infiniswap,Nswap}, can be fitted with machine hot-plugging and hot-removal capabilities to allow applications to scale horizontally. Also, process and virtual migration proposals can be used for vertical scaling. Adopting these approaches, however, either degrades the application’s performance, or involve significant efforts for elasticizing applications that may render them infeasible or undesirable. Clearly, software remains a key problem to achieving elasticity today. \\

This paper addresses the challenge described above. It proposes offering elasticity as a generic service supported by the software system. This service allows applications to scale up, scale out, and scale down to accommodate changing workload needs in an automatic and transparent manner to software developers. This new service is implemented as an integral part of the operating system and allows processes to stretch the limits of available resources beyond one machine, while supporting process mobility between nodes to minimize performance loss stemming from resource distribution.We introduce ElasticOS, a new operating system built on top of Linux, which is a realization of the vision put forward above. Figure 1 provides a high level view of ElasticOS. If each element of the grid represents a node in a datacenter, then we see that an instance of ElasticOS is capable of spanning a
number of nodes in the data center, and that the number of spanned nodes can elastically scale up or down depending upon application demand. This new operating system supports four new primitives: (1) stretching the address space of processes across a cluster of compute nodes, allowing processes to jump (2) from one
machine to another within the set of nodes participating in the stretched address space, to maximize locality, pushing (3) pages between nodes for optimal placement, and pulling (4) memory pages to serve remote page faults. We term our approach of stretching virtual memory address space across multiple physical nodes as
elastic virtual memory. The focus of our efforts towards elasticizing operating systems for cloud applications will be on jointly elasticizing memory and computation. Outside of the scope of our work is elasticizing other elements
of the OS, such as I/O devices and file systems.\\

In the remainder of this disseration, we first describe related work that may help achieve software elasticity in Chapter 2. In that chapter, we also identify what is still missing in the state of the art. Next, in Chapter 3, we first describe our approach to address those gaps. This is followed by a detailed description of
the architecture and components that were introduced into the Linux kernel to achieve ElasticOS. Chapter 4 describes our evaluation of ElasticOS, including analyzing the latency performance versus networked swap, as well as a performance analysis of individual components of ElasticOS. We conclude with a discussion of
future work in Chapter 5.


\fi

\section{ElasticOS Primitives in Action}
\label{sec:abstractions}

In this section, we describe the four primitives through an illustration of a running program.
Figure 2 
graphically presents each of the primitives.
In this figure, we can see nodes 1 and 2, with pages inside of each node -- this represents the physical memory and whether a
given page is used (shaded) or unused (unshaded) in physical memory.
As a starting point, an application is running on a single machine.  Over time, this application
grows in memory use to nearly the size of the amount of memory in the entire node (label 0 in the figure).
This is when ElasticOS decides to stretch the process, that is to scale out by 
using memory on a second node (label 1).  At this point, the memory available to the application has grown
(doubled in the figure, since it is now on two nodes with equal memory, which is not required in ElasticOS).  
ElasticOS can choose to balance the pages at this point, to transfer pages
to the (new) remote node (label 2). These can be chosen by a means, such as least recently used.

Once the process is stretched, this means that the process is effectively running 
on multiple machines, but each node only hosts some of the pages.
At this point, execution continues on the original machine.  As not all
of the pages are on this machine (which would have naturally happened over time, even if we didn't balance pages),
when the process tries to access a page, it might trigger a page fault.  In ElasticOS, the
page fault handler is modified to handle this situation.  At this point, we perform a pull, where a page from 
a remote machine (that caused the fault), is transferred to the local machine and the process is resumed.
The process will be able to make progress, as the page that is being accessed (and caused a fault) is now local.

If space is needed to perform a pull, we can perform a push to free up memory for the incoming page by transferring
a page to a remote node (that we have stretched the application to).  Push
(and pull) is more versatile, as they can be performed proactively as well -- moving pages around, in the background,
to optimize the placement for locality (label 3).

The idea of locality is important, especially in regards to our final primitive, jump.  
Assuming that programs have locality, there is a certain point at which, when we transition
into a new pocket of locality, that the amount of data that forms that locality is high.  It 
is therefore advantageous to jump execution to the data, rather than pull it all into the local node
(as is done in network swap).  In the figure, in steps (4 and 6), the area highlighted in red represents an island of locality that would be more advantageous to jump to rather than pulling the entire group of pages to the local machine.  When to jump is an important decision -- jumping
too much can hurt performance (constantly transferring execution, without making progress), but not jumping 
enough can also hurt performance (transferring lots of data back and forth between machines).  As such,
we created an initial algorithm, and implemented it as a flexible module within which new decision making algorithms
can be integrated seamlessly.

\section{ElasticOS Architecture}

\begin{figure}
	\centerline{\includegraphics[width=\linewidth,keepaspectratio]{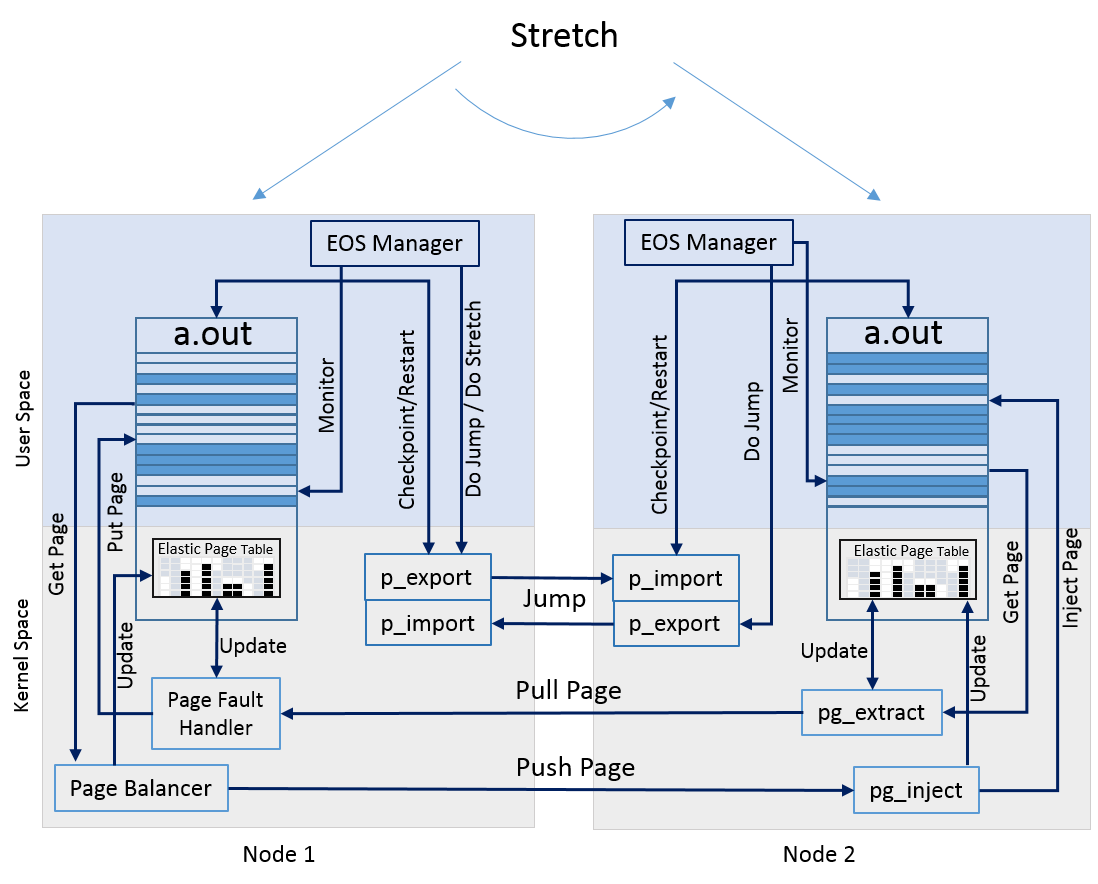}}
	\caption{EOS Architecture.}
	\label{fig:arch}
\end{figure}

In this section, we describe the main components of the ElasticOS architecture.  ElasticOS can be built as a service integrated into existing and commercially-available operating systems. Figure \ref{fig:arch} illustrates the main functional elements that enable a process (e.g., a.out) to be stretched for distributed execution over two ElasticOS nodes.  For clarity purposes, we depict the perspective of pushing and pulling from the perspective of node 1, but in reality all nodes have symmetric capabilities to enable pushing, pulling, and jumping in all directions.

In the remainder  of this section, we will provide a more detailed architectural overview focusing on mechanisms that are roughly OS-independent in order to achieve stretching (\ref{sec:stretching}), pushing (\ref{sec:pushing}), pulling (\ref{sec:pulling}), and jumping (\ref{sec:jumping}). The discussion of OS-dependent elements specific to the Linux implementation is reserved for Section \ref{sec:implementation}.

\subsection{Stretching}
\label{sec:stretching}

\begin{figure}
	\centerline{\includegraphics[width=\linewidth,keepaspectratio]{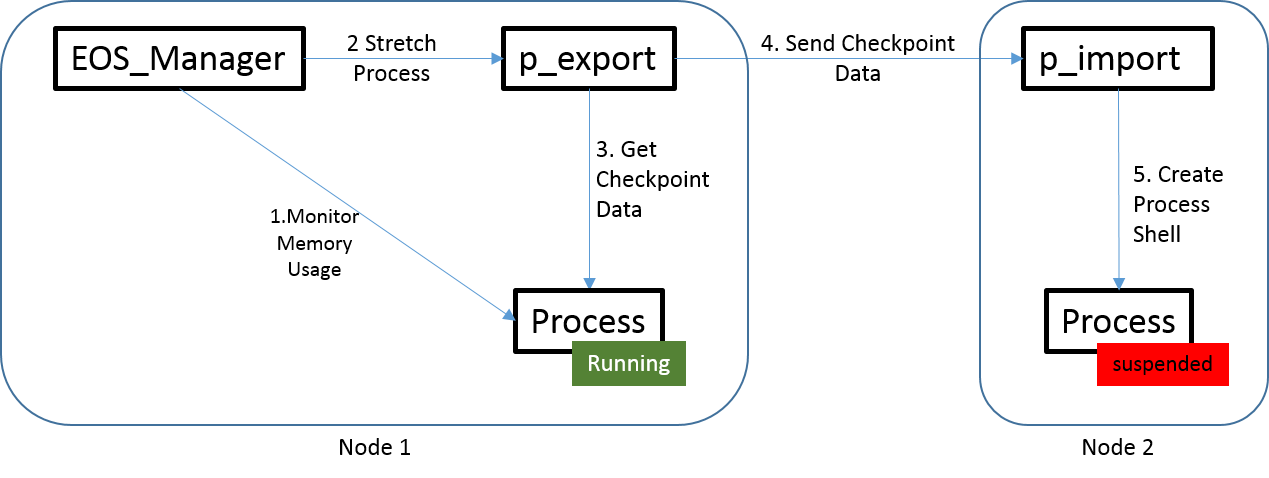}}
	\caption{Stretching.}
	\label{fig:stretching}
\end{figure}

Stretching is responsible for enabling a process to span multiple nodes.  This consists of an initial stretch operation, as well as on going synchronization.

\textbf{Initial stretch operation:}
In order for a process to span multiple nodes, it needs a process shell on each machine.  In this way, 
stretching resembles previous Checkpoint/Restore (C/R) works \cite{CRIU, MOSIX}, except that less information needs to be written into the checkpoint. Here we will need to create a process shell that will remain in a suspended state rather than wholly-independent runnable replica. This makes stretching faster than standard C/R. It requires kernel-space process meta-data. These include virtual memory mappings (mmaps), the file descriptor table, scheduling class, and any other meta-data which is not updated frequently. Other information that is typically modified at a high rate such as pending signals, register state, and stack frames need not be in the checkpoint and will be carried over from the running process whenever it jumps (\ref{sec:jumping}).

As shown in Figure~\ref{fig:stretching}, stretching is triggered by the EOS manager, which continuously monitors process' memory usage and issues a newly-created signal (\texttt{SIGSTRETCH}) whenever it detects a process that is too big to fit into the node where it is running. Our special kernel-space handler (eos\_sig\_handler) intercepts the signal and instructs the process-export module (p\_export) to send the checkpoint using a pre-created TCP socket to a process-import module (p\_import) waiting in the other node. The latter will, then, create a shell process by allocating the necessary kernel-space structures and filling them in with checkpoint data.

\textbf{State Synchronization:}  After the process has been stretched, and its replica has been created on another machine, additional changes in process state on the first machine will need to be propagated to the replica.  This is handled in two ways.  Rapid changes in state are handled using the jumping mechanism, as explained later.  Changes in state at a more intermediate time scale such as mapping new memory regions and opening or closing files are handled using multicast sockets to listeners on each participating node.

One of the pitfalls to avoid here is that the operating system scheduler may delay flushing all such synchronization messages until after a jump is performed. If this happens, the system may arrive at an incorrect state or even crash. So, it is crucial to flush all synchronization message before a jump is performed. 

\subsection{Pushing}
\label{sec:pushing}

\begin{figure}
	\centerline{\includegraphics[width=\linewidth,keepaspectratio]{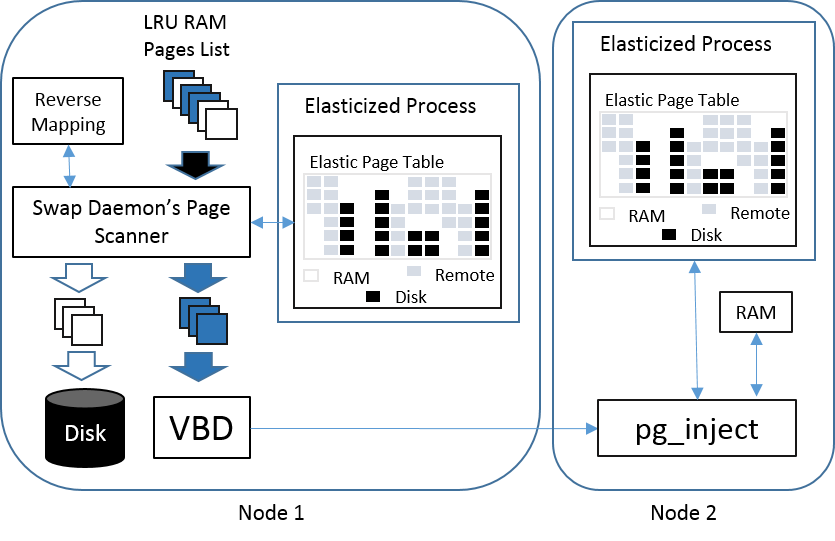}}
	\caption{Pushing.}
	\label{fig:pushing}
\end{figure}

Now that the process has presence on more than one machine, its memory pages are \textit{pushed} between nodes in order to balance the load among participating nodes. Our page pusher piggybacks on existing OS's swap management (See Figure \ref{fig:pushing}).

Typically, the swap daemon scans least-recently used (LRU) lists to select least recently used page frames for swapping. Our page balancer modifies this page scanner in order to identify pages mapped by elasticized processes (shaded pages in Figure \ref{fig:pushing}) using reverse mapping information associated with the page. These are then sent to a virtual block device client (VBD), similar to the one described in \cite{Infiniswap}, after updating the respective page table entries (PTEs) in the elastic page table. The VBD then forwards the page along with relevant information such as process ID, and the page's virtual starting address to the page injection module (pg\_inject) on the node, which will then allocate a new page, fill it with the proper content, and update the replicas elastic page table. 

Maintaining accurate information in the elastic page tables when pushing pages is very crucial to correct execution. As we will see later, jumping depends on this information for locating pages in the system.

\subsection{Pulling}
\label{sec:pulling}

\begin{figure}
	\centerline{\includegraphics[width=\linewidth,keepaspectratio]{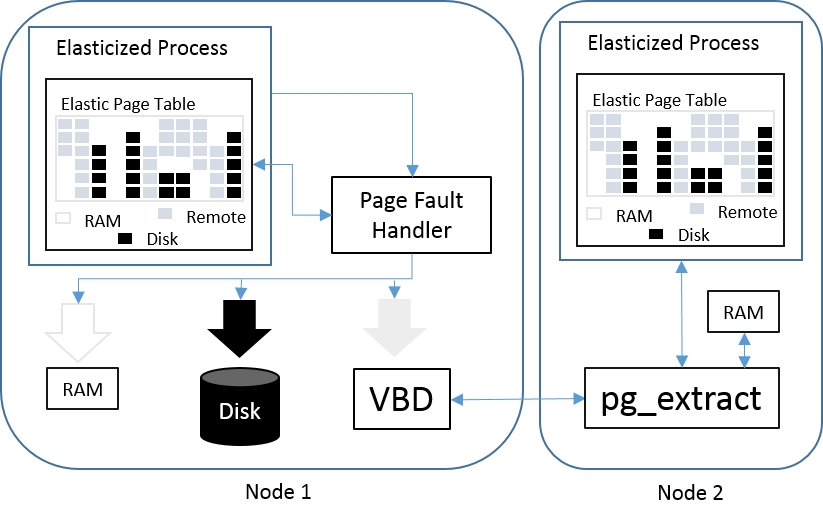}}
	\caption{Pulling.}
	\label{fig:pulling}
\end{figure}

Partitioning the process's memory footprint will, inevitably, result in references to remote pages. These are handled by our modified page fault handler (Figure \ref{fig:pulling}). On a page fault, the handler will consult the elastic page table to identify the page's location. If it happened to be on a remote node, the page's starting virtual address and process ID is forwarded to the VBD, which will then contact the page extraction module (pg\_extract) on the respective node to pull the page. Once it receives the page's content, the VBD client, then restores the process's access to the page.

Whenever a remote page fault is handled as described above, page fault counters are updated. This is required by ElasticOS's jumping policy (Section \ref{sec:jumping}), which will always try to co-locate execution with its most-referenced memory.

\subsection{Jumping}
\label{sec:jumping}


\begin{figure}
	\centerline{\includegraphics[width=\linewidth,keepaspectratio]{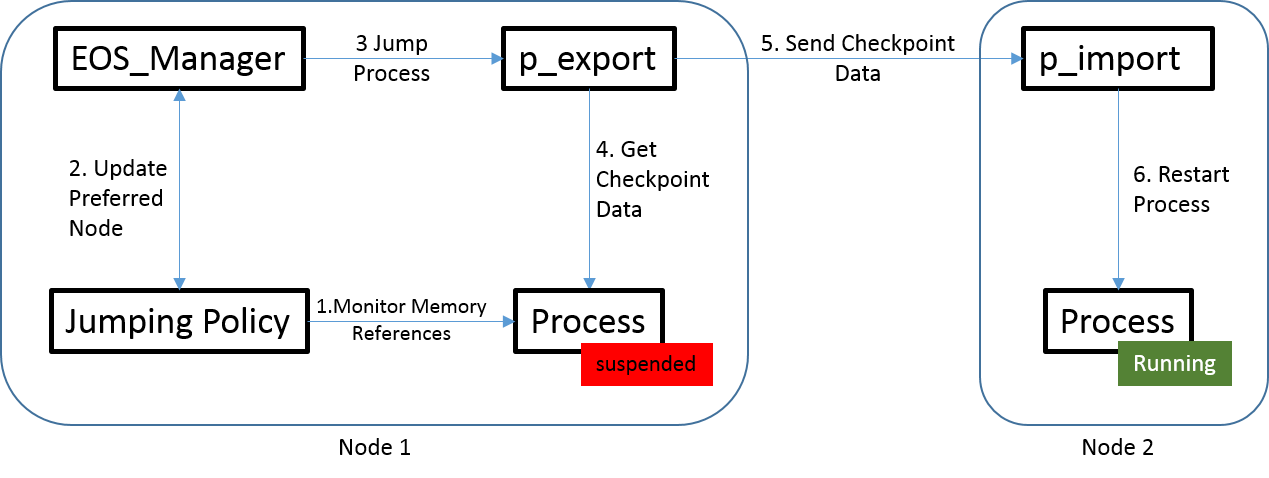}}
	\caption{Jumping.}
	\label{fig:jumping}
\end{figure}

Jumping is the act of transferring execution from one node to another. For this, there is both a jumping mechanism that performs
a lightweight process migration, and the jumping policy to determine when to jump.

\textbf{Jumping mechanism:}
Jumping is an lightweight mechanism similar to checkpoint/restore systems.  In contrast to stretching, with jumping, the process does actually transfer execution, and only carries in the checkpoint the information that changes at a high rate. This includes CPU state, the top stack frames, pending signals, auditing information, and I/O context. The overall size of jumping checkpoint data is dominated by the stack frames, so it is very important to include only the topmost stack memory pages that are necessary for correct execution. 

As shown in Figure~\ref{fig:jumping}, whenever a jump is deemed necessary by the jumping policy in the EOS Manager, it sends a special signal (\texttt{SIGJUMP}) to the process, which is then routed to the eos\_sig\_handler which will then instruct the p\_export module to checkpoint the process and send the information to the other node's p\_import module. The latter will fill in the appropriate kernel-space structures and set the process's state to runnable. Notice here that when jumping, no new structures need to be allocated since the process has been already stretched to the target node. Also, notice that the process at the source node will remain in a suspended state.  In essence, jumping resembles rescheduling a process from one CPU to another across the boundaries of a single machine.

\textbf{Jumping Policy Algorithm:}  Maximizing locality is crucially important to the application's performance. A naive approach to moving execution and memory pages around in the system will, inevitably, increase the rate of remote page faults leading to poor performance. Thus, a good policy for moving processes close to their most frequently used memory is of critical importance.
ElasticOS can achieve this goal by overcoming two challenges, namely having a good sense of how to group inter-dependent memory pages together on the same node, and detecting which of those groups is the most frequently accessed one.

The first challenge can be overcome by taking advantage of the natural groupings memory pages belonging to an application tend to form due to recency of reference. This property is already evident in the wide adoption of the LRU algorithm for page replacement in most modern OSs. Thus, we can extend LRU algorithms to work in a multi-node system, where pages evicted from one node's RAM are immediately shipped to another node via our pushing mechanism. 

The second challenge can be addressed by implementing a jumping policy that: 1) monitors the process's page accesses to find the "preferred" node, and 2) reschedules the process to the preferred node if it is running on any of the other ones.

Bear in mind that accurately tracking memory references for a particular process can be a challenging task since CPUs do not report every memory access to the OS for performance reasons. This leaves us with options that provide the "next best thing", such as counting the number of time the CPU sets \texttt{PG\_ACCESSED} flag for a particular page frame when it is accessed in the X86\_64 architecture or tracking handled page faults.

\section{ElasticOS Implementation}
\label{sec:implementation}

We implemented ElasticOS as a fork of the Linux kernel v2.6.38.8.  We chose the 2.6 kernel because it contains
the key features that we needed to demonstrate elasticity, e.g. support for 64-bit x86 architectures and a reasonably featured virtual memory manager, while avoiding unnecessary complexity and instability in later kernels.  

\textbf{System Startup:}
\label{sec:discovery}
 Whenever a machine starts, it sends a message on a pre-configured port announcing its readiness to share its resources. The message includes two groups of information. First, connectivity parameters such as IP addresses and port numbers. Second, information noting the machine��'s available resources, which includes total and free RAM. 
 Next, each participating node records the information received about the newly-available node and initiates network connections for the various clients.
 Finally, EOS manager is started, which will periodically scan processes and examines their memory usage searching for opportunities for elasticity.  

Identifying such opportunities can be achieved by examining the per-process counters Linux maintains to keep track of memory usage. They include: 1) \verb|task_size| inside each process's memory descriptor (i.e., \verb|struct mm_struct|) which keeps track of the size of mapped virtual memory, 2) \verb|total_vm| inside the same structure to track the process's mapped RAM pages, 3) \verb|rss_stat| of type \verb|struct mm_rss_stat| which contains an array of counters that further breaks down \verb|task_size| into different categories (i.e., anonymous and file-mapped RAM pages used, and swap entries), and 4) \verb|maj_flt| variable inside the \verb|struct task_struct| which counts the number of swap-ins triggered by the process.

Linux also maintains memory utilization indicators called watermarks. There are three levels of watermarks:
min, low, and high. These levels drive the kernel swap daemon's (kswapd) activity. When memory usage
reaches the high watermark, page reclaim starts, and when it goes down to low watermark, page reclaim
stops. 

ElasticOS leverages these watermarks and the level of kswapd's activity to detect periods of memory pressure. Further, it identifies specific memory-intensive processes using the counters mentioned above and marks them for elasticity. 

\textbf{Stretching Implementation:}
\label{sec:stretching-impl}
The Linux kernel forces each process to handle pending signals upon entering the CPU. This is when our in-kernel signal handler, the p\_export module, checks for pending ElasticOS-specific signals. This design choice of checkpoint creation logic placement gives us access to register state, while preventing the process from updating its own memory while a checkpoint is in progress.

The handler, then, accesses the process information while writing them to a socket initialized during system startup. At the other end of the socket, the p\_import module collects the information and uses it to create the new shell process. 

The key items that are included in this checkpoint consist of: contents of the process descriptor (\texttt{struct task\_struct}), memory descriptor and (\texttt{struct mm\_struct}) virtual memory mappings (\texttt{struct vm\_area\_struct}), open files information (\texttt{struct files\_struct}), scheduling class information (\texttt{struct sched\_class}), signal handling information (\texttt{struct sighand\_struct}), and few others. The overall size of the this checkpoint in our experiments averages around nine kilobytes, which are dominated by the size of the process's data segment which is also included in the checkpoint. 

Note, that we do not need to copy memory pages containing the code, since our implementation assumes that the same file system is available on all participating nodes. Instead, we carry over with the checkpoint data the mapped file names. Our p\_import  module will locate and map the same files at the appropriate starting addresses. 

P\_import handle the process creation the same way as if it were forked locally while substituting missing values with others from the local machine. For example, it assigns the newly created process a "baby sitter" to replace the real parent from the home node.

Finally, the p\_import module leaves the newly created process in a suspended state and informs the p\_export module that it can allow the original process in the source node to resume execution.

\textbf{Pushing and Pulling Implementation:}
We extend Linux's second-chance LRU page replacement algorithm by adding multi-node page distribution awareness to it. In this version, pages selected for swapping out belong to elasticized processes and are pushed to another node and injected into the address space of the process’
duplicate there. 
Second-chance LRU groups pages in reference-based chronological order within the pages list. So, it is most likely that pages at
the rear of the queue, which are typically considered for eviction, are related in terms of locality of reference.

One challenge that needed to be solved to implement page balancing is identifying pages belonging to
an elasticized process and what virtual address they are mapped to. Luckily, Linux maintains a functionality
called reverse mapping, which links anonymous pages to their respective virtual area map. By walking this
chain of pointers and then finding which process owns that map, we can tell them apart from other pages
owned by other processes in the system. Then, with simple calculations we can find the starting virtual
address of that page.
As for moving pages from one machine to another, we created a virtual block device (VBD) that
sends page contents using a socket connected to a page server on the other machine (VBD Server) rather
than storing it to a storage medium. This was shown in Figure \ref{fig:pulling}. This virtual block device is added to
the system as a swap device. All pages belonging to an elasticized process sent to the other machine are
allocated swap entries from this device. This swap entry is inserted into the page table of the elasticized
process where the page is mapped. As a result, if that page needs to be faulted in later on, the swap entry
will route the page fault to our VBD. This design choice allows us to reuse Linux's page eviction and faulting
code.


\textbf{Jumping}:
Whenever a remote page fault occurs, a remote page counter is incremented. We keep track of the number of remote page faults to use it later on for jumping.  As the page remote fault counter builds up, it will show the tendency of where page faults are "going". If the remote faults count value hits a predetermined threshold, then the system could 
determine that the process would better exploit locality of reference if it jumps to the remote node. 
Jumping starts by sending a special signal to the target process, which is handled by an in-kernel
checkpoint module. This module will, then, copy only the necessary information for the process to resume on the other node. This information includes: 1) the thread context, which contains the register state and other flags, 2) pending signals (i.e., \texttt{struct sigpending} contents inside \texttt{struct task\_struct}), 3) auditing counters, and 4) the stack page frames (i.e., RAM pages mapped by the \texttt{vm\_area\_struct} with the flag \texttt{VM\_GROWSDOWN} set). In our tests, the checkpoint size was roughly 9KBs and was dominated by the two stack page frames (4KBs each). Other information about the process will be synchronized using a special module described next. These pieces of information are sent to the restart module in the 
remote node via a pre-established TCP connection. In its turn, the restart module updates the process information with the checkpoint data, and sends a (\texttt{SIGCONT}) to the process. This will inform the scheduler that it is ready to run again. The process on the source machine will remain in an interruptible wait state (i.e., suspended). This will guarantee that only one clone of the process is running at any given instance.

\textbf{State Synchronization}: 
The state synchronization component is built as a collection of user-space programs and a kernel module. The user space portion simply sets up the network connections and then passes their socket descriptors to the kernel module, which exposes hook functions to the main kernel. 

When an elasticized process issues a system call that modifies its in-kernel data structures (e.g., \texttt{mmap}), the appropriate kernel module hook function is called (e.g., \texttt{sync\_new\_mmap}), which will then multi-cast a message to all participating nodes. The message will contain all necessary information (e.g., region's starting address, its length, mapping flags, and file name) to apply the same operation on all process replicas. Multi-cast listeners, then, relay the message to the appropriate hook functions, who will apply the change (i.e., call \texttt{mmap} on the process replica).

\section{Performance Evaluation}
In this section, we focus on evaluation of the performance of ElasticOS.  Specifically, we look
to quantify the benefit of joint disaggregation (memory and computation) by comparing against network swap, which is
a one dimensional (scaling memory), which has previously been shown to have performance benefits over
not scaling memory~\cite{Nswap, Infiniswap}.  We note that we do not explicitly compare against just process
migration, as the use cases are different, where process/VM migration is commonly used to move execution
permanently and triggered by contention for resources or for other 
operational reasons (e.g., planned maintenance) -- making it heavier weight and not well suited for comparison.


\subsection{Experimental Setup}

We evaluated ElasticOS on the Emulab testbed~\cite{emulab}. We used Emulab D710 nodes with 64-bit Quad Core Xeon processor, 12 gigabytes RAM, and a gigabit NIC.  We choose Emulab D710 nodes because they support Linux kernel 2.6. Our experimental setup for each experiment consists of two nodes connected via gigabit Ethernet ports, transported through a network switch.

To evaluate, we ran tests on a variety of algorithms representing the type of processing 
that would be a target use case for ElasticOS -- large graphs or lists to be processed. 
Shown in Table~\ref{tab:alg} is a summary of these applications, and the footprint of each 
application -- note that the footprint goes beyond the limits of a single node in Emulab.
Specifically, these algorithms typically use 11GB of memory on the first machine, and stretch to 
a remote machine for the additional memory.

\begin{table}
\centering
\caption{Tested algorithms and their memory footprints.}
\label{tab:alg}
\begin{tabular}{|l|c|}
\hline
Algorithm          & Memory Footprint                \\ \hline
Depth First Search & 330 million nodes (15 GB)       \\ \hline
Linear Search      & 2 billion long int (15 GB)      \\ \hline
Dijkstra           & 3.5 billion int weights (14 GB) \\ \hline
Block Sort         & 1.8 billion long int (13 GB)    \\ \hline
Heap Sort          & 1.8 billion long int (14 GB)    \\ \hline
Count Sort         & 1.8 billion long int (14 GB)    \\ \hline
\end{tabular}
\end{table}

In our experimental setup, we employed a basic jumping algorithm to trigger transfer of execution.  A simple remote page fault counter is updated for each remote pull, and whenever a counter threshold value is reached, then a process will jump its execution to the remote machine.  In addition, the counter is then reset.  We tested the algorithms with different counter threshold values (32 up to 4M). 

For each algorithm, we measure its execution time as well as network traffic generated, and compare results of ElasticOS and network swap.  To provide a comparison with network swap, hereafter termed Nswap, in a manner which isolated the gains to simply the benefit of jumping and not any implementation differences, we use ElasticOS code, but disable jumping. In this way, Nswap tests pin a process on one machine, but use the memory of a remote machine as a swap space. In our experiments, both ElasticOS and Nswap spanned two machines.  Emulab provides isolation of networking and execution in the testbed from external disturbances.
\begin{table}
	\centering
	\caption{Micro-benchmarks of ElasticOS primitives.}
	\label{tab:prim}
	\begin{tabular}{|l|l|l|}
		\hline
		Primitive & Latency     & Network Transfer \\ 
		\hline \hline
		Stretch   &  2.2ms      & 9KB \\ 
		\hline
		Push      &  30-35us     & 4KB \\ 
		\hline
		Pull      & 30-35us      & 4KB \\ 
		\hline
		Jump      & 45-55us      & 9KB \\ 
		\hline
	\end{tabular}
\end{table}

\subsection{Micro-benchmarks}

An important metric when evaluating ElasticOS is the performance of each individual primitive.  These are summarized in Table~\ref{tab:prim}, based
on our measurements on Emulab D710 nodes.  We'll note that jumping is very fast, taking only 45-55 microseconds.  This is
substantially lower than reported numbers for process or VM migration, which are measured in seconds (e.g., one benchmark states
CRIU's downtime is roughly 3 seconds~\cite{criu_perf}).  Stretching is
also only performed once -- when a decision is made that this process would benefit from scaling in the future.

We measured pushing and pulling to be between 30-35 microseconds -- roughly the time to submit the request and transfer a page (4KB) of
data across a network.  
For jumping to be effective in speeding up execution in ElasticOS, there must be locality.  That is, the time for a single jump must 
be less than the time for the number of remote page pulls that would be saved by jumping.  
For our performance microbenchmarks, 
for jumping to be efficient, the process should save at least two remote page pulls.  As we show next, the locality is much greater than
this, resulting in substantial speedups.


\begin{figure}[ht] 
	\centering
	\includegraphics[scale=0.65,trim={0 1.5cm 0 1.5cm}, clip]{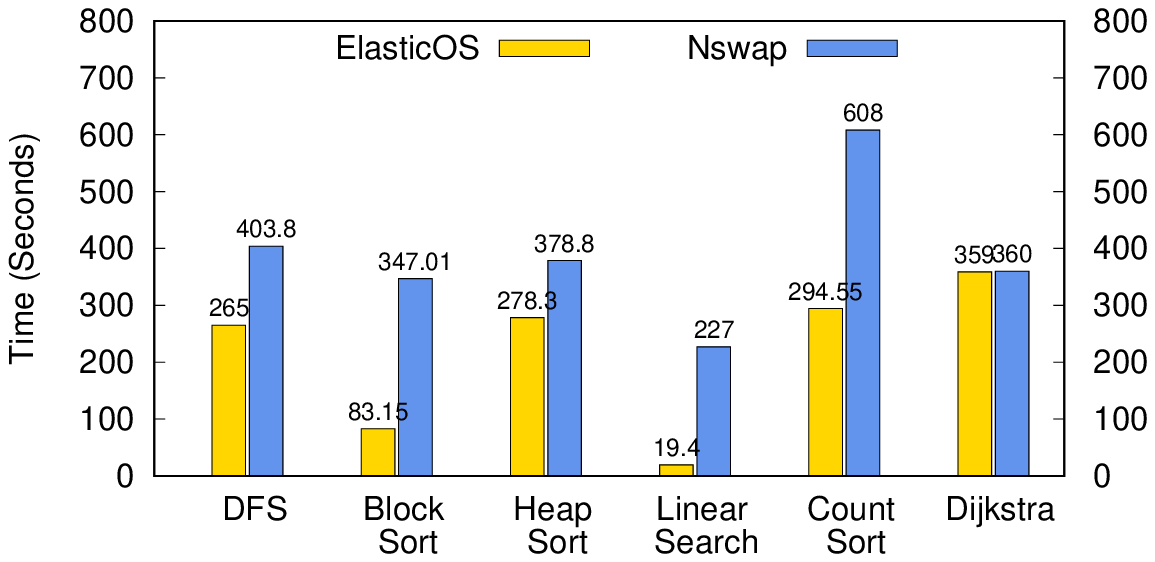}
	\caption{Execution Time Comparison.}
	\label{fig:execution}
\end{figure}

\subsection{Execution Time and Network Traffic}

There are two key metrics to consider when comparing ElasticOS (with jumping, pulling and pushing), to network swap (just pulling and pushing).  The first is overall execution time.
Here, the key premise behind jumping is that to exploit locality, we should transfer execution to where the data is, rather than pull in the data 
to where the execution is.  The second is the amount of network traffic -- jumping needs to transfer context (e.g., the current stack), and pulling/pushing 
transfers pages.  

In Figure~\ref{fig:execution}, we show our measured average execution time for both Nswap and ElasticOS for each of the algorithms
we have evaluated.  These execution times are averaged over four runs using the threshold that achieves the most improvement.  We observe that in the best case, ElasticOS shows substantial performance benefits for most algorithms.  For example, Linear Search experienced about an order of magnitude speedup in execution performance, Depth First Search (DFS) achieved about 1.5X delay improvement, while Dijkstra's algorithm achieved no speedup.  

Table \ref{tab:jumping} describes the specific threshold values where best performance was achieved in ElasticOS for each algorithm.  It also lists the total number of jumps at that threshold as well as the frequency of jumping for each algorithm at that threshold.  The jumping rate ranges from less than once per second to hundreds of times per second.

While Figure~\ref{fig:execution} represents the best case, we were also interested in understanding whether we could find universal threshold values that achieves performance improvements - perhaps not the best - regardless of the algorithm.  Our analysis found that, regardless of the algorithm, using any threshold value above 128, Elastic OS performs better than Nswap for any algorithm, either in delay, network overhead or both.  

\begin{figure}[ht] 
	\includegraphics[scale=0.65,trim={0 1.2cm 0 1.5cm}, clip]{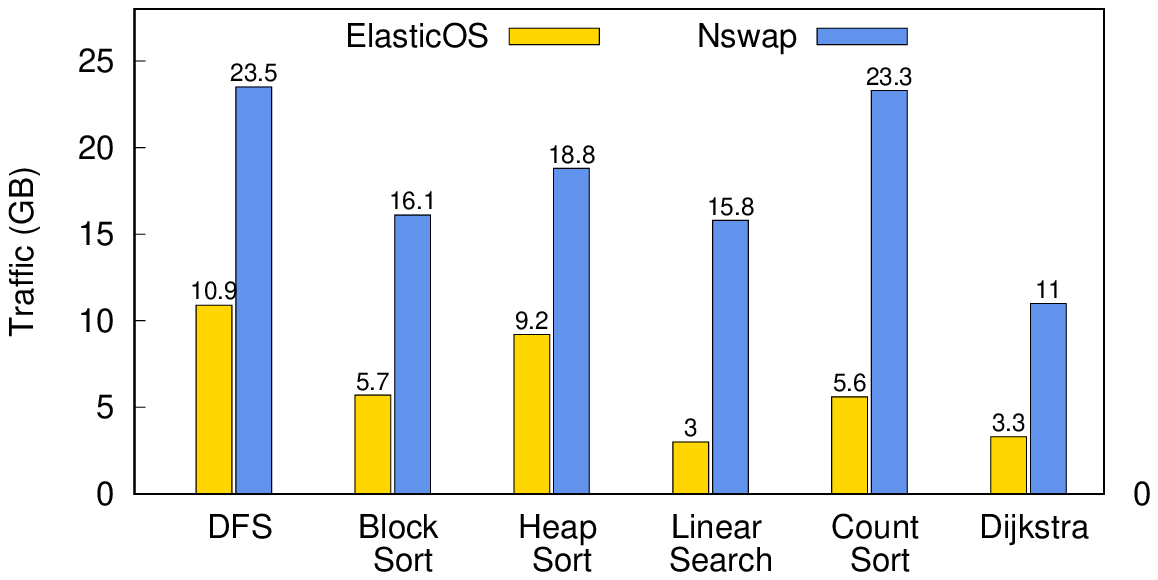}
	\caption{Network Traffic Comparison.}
	\label{fig:traffic}
\end{figure}

The use of jumping to exploit locality improves the execution time by enabling more local pages to be accessed, rather than
needing to go across a network (which is orders of magnitude slower).  This also reduces the amount of network traffic, even taking
into account the data transfer needed to perform a jump.  Shown in Figure \ref{fig:traffic} are our measured results for each of the algorithms
tested.  We can see that ElasticOS reduces the amount of traffic on the network for all algorithms tested by a significant amount -- from a 5x reduction 
for Linear Search to about 2x reduction for DFS.  By avoiding the process of swapping in and out to remote machines through lightweight jumping, 
we save a large amount of data and control traffic associated with avoidable remote page faults.  Also, even if we did not achieve any delay improvements 
running ElasticOS, we still can obtain network traffic reduction.  For example, Dijkstra's algorithm did not achieve any delay improvement, even though Table~\ref{tab:jumping} shows that Dijkstra had 520 jumps, but these jumps helped reducing its network overhead by 70\%.  
In examining the behavior of Dijsktra's, its initial set of jumps before settling down to execution on one machine resulted in substantial overhead savings.  
\begin{table}
	\centering
	\caption{Jumping Thresholds.}
	\label{tab:jumping}
	\begin{tabular}{|l|l|l|l|}
		\hline
		Algorithm & Threshold   &  Number  & Jumping \\ 
		          &             &     of jumps & frequency \\
			        &			&     & (jumps/sec)\\
		\hline \hline
		DFS   &  8K      & 180 &  0.6 \\ 
		\hline
		Block Sort      &  512     & 1032 & 12.3 \\ 
		\hline
		Heap Sort      & 512      & 3454 & 12.4 \\ 
		\hline
		Linear Search      & 32      & 3054& 157.4\\ 
		\hline
		Count Sort      & 4096      & 198& 0.6\\ 
		\hline
		 Dijkstra      & 512      & 520& 1.4\\ 
		\hline
	\end{tabular}
\end{table}
\subsection{Understanding Application Specific Behavior}

We previously showed that each algorithm has a varying degree of improvements.  While the simple
reasoning is that it is due to locality, here we examine three of the algorithms in detail to 
really understand this behavior.

\subsubsection{Linear Search}
\begin{figure}[ht]
	\includegraphics[scale=0.6,trim={0 1.5cm 0 1.5cm}]{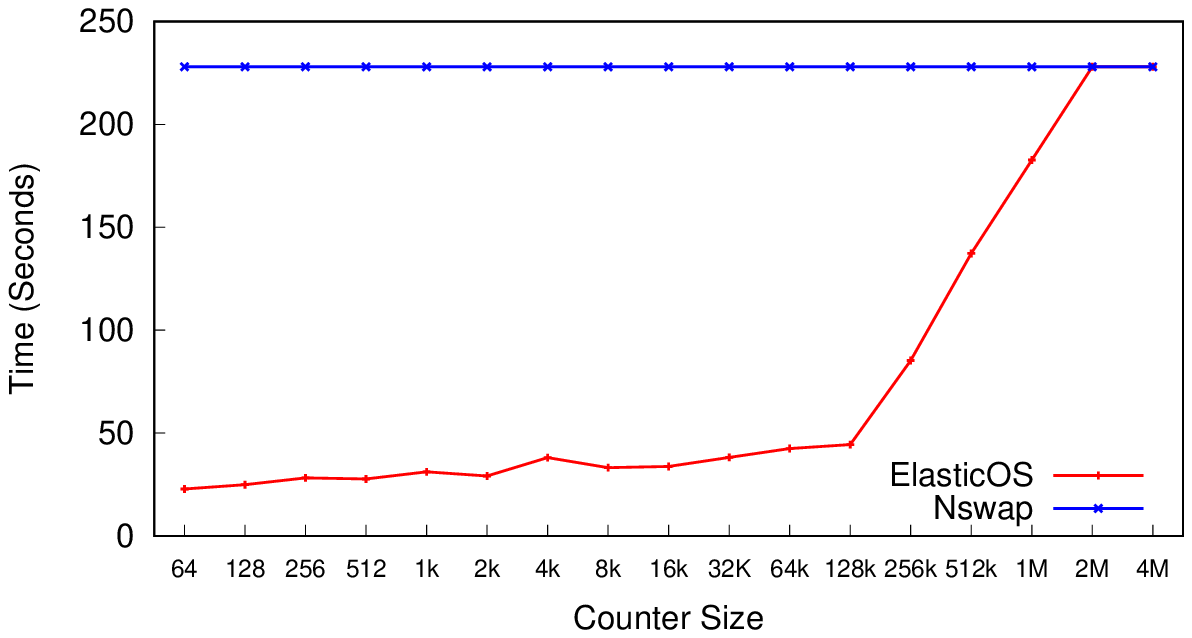}
	\caption{Linear Search Execution Time.}
	\label{fig:lienar_time_ext}
\end{figure}

For Linear Search, the memory access pattern is simple and predictable, namely the memory address space is accessed in a linear fashion.  As a result, consecutive memory pages tend to age in LRU lists together, and end up being swapped to the remote machine together. When a process jumps towards a remote page, it is very likely for the process to find a chunk of consecutive pages to access, exploiting locality of these pages, which saves the process a significant amount of time by avoiding swap overhead. Figure \ref{fig:lienar_time_ext} shows delay improvements on Linear Search with respect to jumping threshold.  Linear Search tends to perform better when the counter threshold is smaller, hence jumping early is better when accessing the address space in a linear fashion. Table~\ref{tab:jumping} shows the highest frequency of jumping for linear search, as well as the lowest threshold value used. We also observe that as the threshold for jumping increases, jumping will occur less often and eventually not at all, hence explaining why the delay curve for ElasticOS converges to Nswap.

\subsubsection{Depth First Search}
\begin{figure} 
	\includegraphics[scale=0.6,trim={0 1.5cm 0 1.5cm}]{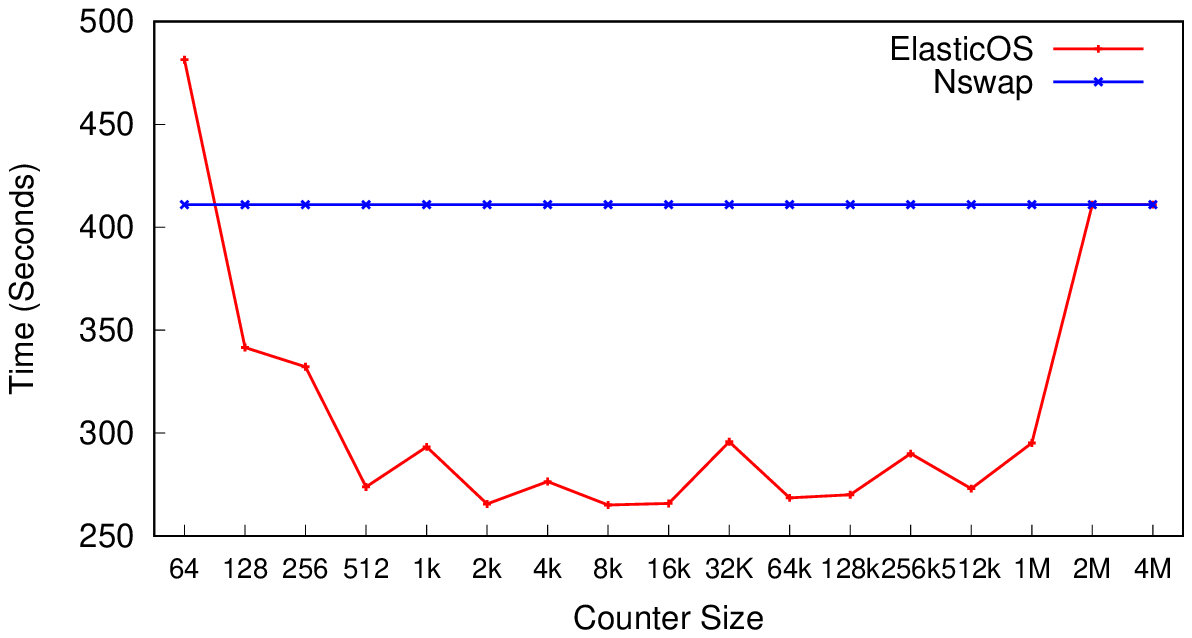}
	\caption{Depth First Search Execution Time.}
	\label{fig:dfs_time_ext}
\end{figure}

\begin{figure}[ht]
	\includegraphics[scale=0.6, trim={0 1.5cm 0 1.5cm}]{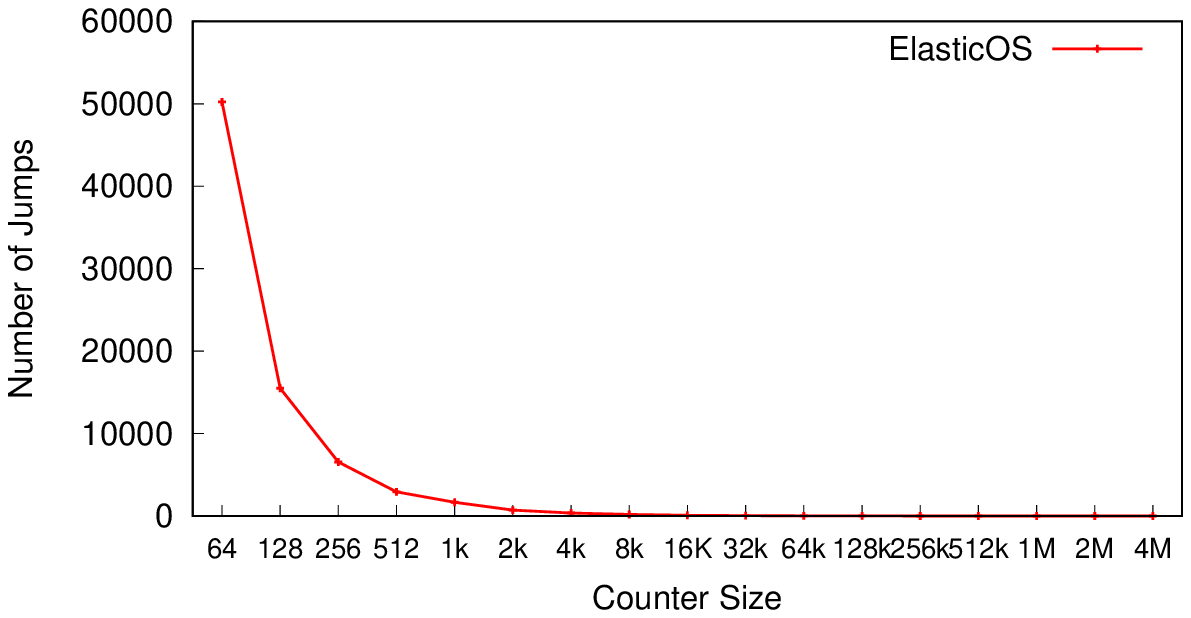}
	\caption{Depth First Search  Number of Jumps.}
	\label{fig:dfs_jumps_threshold}
\end{figure}

On the other hand, Depth First Search has a non linear memory access pattern.  When the algorithm starts a depth first search, the search starts at the root node, and traverses the graph branch by branch, from root to the end (depth) of the branch.  While the graph nodes are laid out in a certain order in the memory space, 
the access pattern of DFS does not match this layout.  This increased randomness of access to pages means that there is less locality to exploit on each jump than occurred for Linear Search, and hence less gain versus Nswap compared to Linear Search.  
Figure \ref{fig:dfs_time_ext} shows different execution times of DFS for various counter threshold sizes.  ElasticOS achieves at best about a 1.5x improvement in delay over Nswap across a wide range of counter thresholds, namely larger than 64.  However, for very small values of threshold less than or equal to 64, DFS performs worse.  Figure~\ref{fig:dfs_jumps_threshold} shows that when the threshold value is very small, DFS experiences a large number of jumps. Also, our tests showed that DFS's best performance happens when the threshold value is large compared to other algorithms as shown in Table~\ref{tab:jumping}.

\begin{figure} 
	\includegraphics[scale=0.6,trim={0 1.5cm 0 1.5cm}]{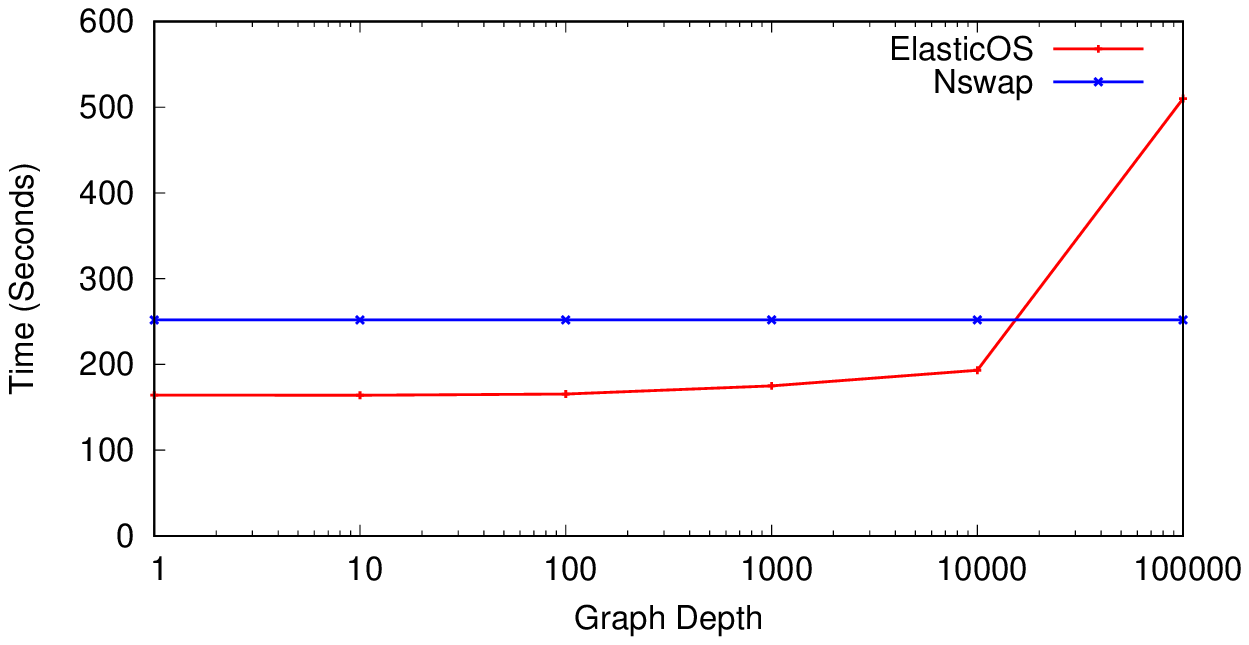}
	\caption{Depth First Search Performance on Different Depths.}
	\label{fig:dfs_time_depths}
\end{figure}

\begin{figure} 
	\includegraphics[scale=0.6,trim={0 1.5cm 0 1.5cm}]{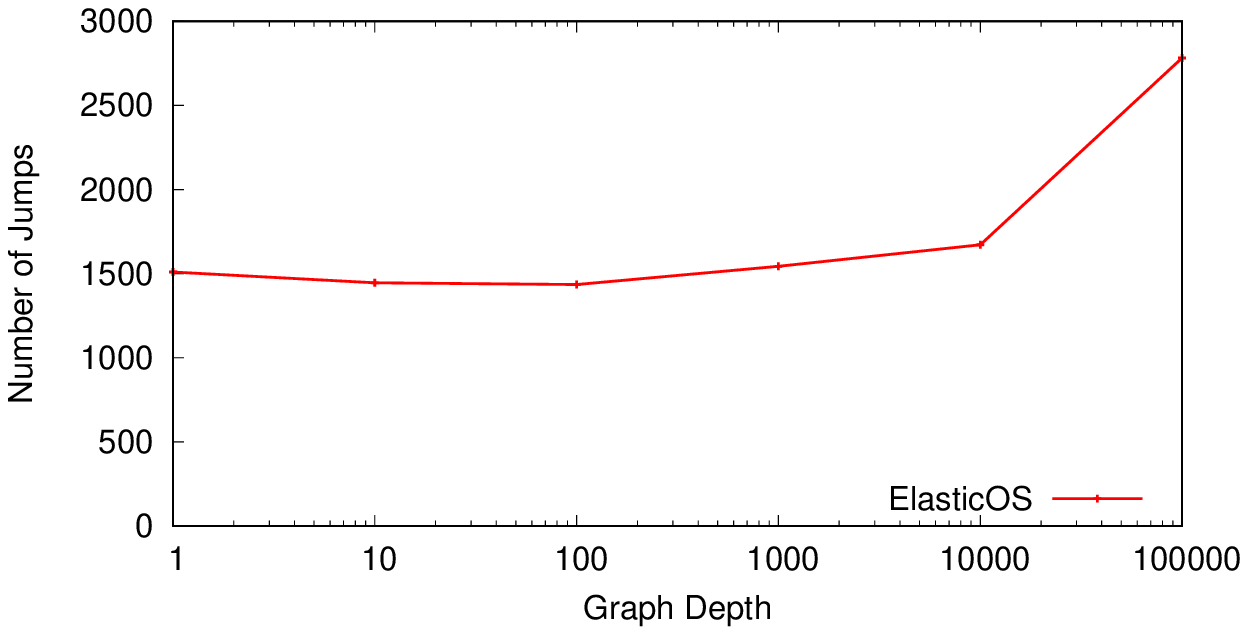}
	\caption{Depth First Search Jumps on Different Depths.}
	\label{fig:dfs_jumps_depths}
\end{figure}

The shape of the graph in DFS can also impact the memory access pattern.  For example increasing the depth of the graph would make branches longer, resulting in a longer branch that occupies more memory pages, increasing the chance of a single branch having pages located both on local and remote machines.  This would increase the chances of jumping more and performing poorly.  Figure \ref{fig:dfs_time_depths} shows DFS performance on ElasticOS for different graph depths with a fixed jumping counter size of 512. Increasing the graph depth eventually results in poorer performance.  Figure \ref{fig:dfs_jumps_depths} shows that this poorer performance occurs when there is excessive jumping for deep graphs.  To make ElasticOS perform better on such graph depth we need to increase the jumping counter size to values larger than 512, to avoid jumping too much.


\subsubsection{Dijkstra's Algorithm}
\begin{figure} 
 	\includegraphics[scale=0.6,trim={0 1.5cm 0 1.5cm}]{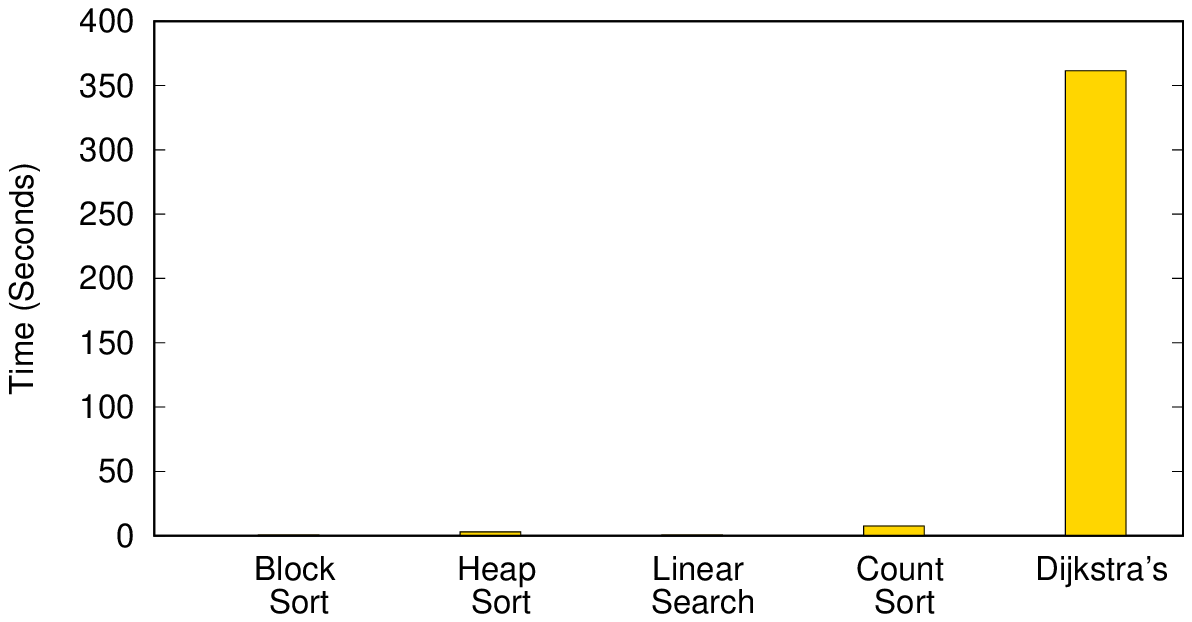}
 	\caption{Maximum Time Spent on a Machine without Jumping.}
 	\label{fig:max_time_no_jump}
 \end{figure}

ElasticOS achieved very little gain when executing Dijkstra's algorithm when compared to Nswap.  
Dijkstra's algorithm scans through an adjacency matrix, then learns and stores information about the shortest path in a separate array.  However, Dijkstra does not necessarily access all nodes in the adjacency matrix, because some nodes are not connected, or one of the paths was excluded for being too long. Since Dijkstra's algorithm keeps track of the shortest path nodes in a separate array, it only accesses the adjacency matrix nodes once, and keeps useful information in the shortest path array.  Based on how Dijkstra's algorithm works, it does not access memory frequently, and only accesses part of the allocated memory.  Therefore, most of Dijkstra's execution time does not involve many remote page faults. Since jumping saves time wasted on remote page faults, Dijkstra does not gain much delay improvement, because it does not jump due to very small number of remote page faults. Figure \ref{fig:max_time_no_jump} confirms that Dijkstra's algorithm spends most of its execution time on one machine without jumping.  Our experiments showed that only a relatively small set of jumps happened at the beginning, and the rest of the time execution stayed on one machine.

\section{Discussion and Future Work}
We intend to upgrade ElasticOS to a newer version of Linux. 
We plan to investigate improved jumping algorithms that better exploit locality by actively learning about elasticized process' memory access patterns during run time and employing adaptive jumping thresholds.  Probabilistic models will be investigated.  In addition, we will explore whether incorporating into the jumping decision the burstiness of remote page faulting brings any benefit.
Also, we are considering a more proactive approach to controlling the swap out operation for elasticized processes by modifying kswapd. If we selectively swap out pages to remote machines, we might be able to create islands of locality on remote machines,  thus, making jumping more efficient. We also can pin memory pages, and prevent them from being swapped, which would allow us to control how the memory address space is distributed across participating machines.
We plan to test a wider variety of algorithms, including SQL-like database operations.  We intend to expand testing to more than two nodes.



\section{Conclusion}
In this paper, we have implemented within Linux four new primitives, namely stretch, push, pull, and jump, to support scaling as an OS abstraction.  This extended Linux system is called ElasticOS.  These primitives transparently achieve joint disaggegration of computation and memory, enabling both data to move towards execution, as well as execution to move towards data within a stretched address space spanning multiple nodes in a data center.  Our evaluation results were obtained from Emulab deployments of ElasticOS testing a variety of different application algorithms, and indicate that such joint disaggregation achieves up to 10X speedup in execution time over network swap, as well as 2-5X reductions in network overhead.

\section{Acknowledgments}

This research was supported by NSF CCF grant \# 1337399, funded under the NSF program Exploiting Parallelism and Scalability "XPS: SDA: Elasticizing the Linux Operating System for the Cloud".  We also wish to thank Sepideh Goodarzy and Ethan Hanner.

\newpage
{\footnotesize \bibliographystyle{acm}
\bibliography{bibliography}}


\end{document}